\documentclass[referee]{aa}

\usepackage{amsmath}
\usepackage{amssymb}
\usepackage{latexsym}
\usepackage{subfig,graphicx}
\usepackage{txfonts}
\usepackage{natbib}
\bibpunct{(}{)}{,}{}{,}{,}
\usepackage{color}
\usepackage[mathcal]{euscript} %Allows order of symbol \mathcal{O}

\begin{document}
\title{Observations of quasi-periodic phenomena associated with a large blowout solar jet}

\author{ R. J. Morton$^1$, A. K. Srivastava$^{2,}$$^1$, \& R. Erd\'{e}lyi$^1$}

\institute{$^1$Solar Physics and Space Plasma Research Centre
(SP$^2$RC), University of Sheffield, Hicks Building, Hounsfield
Road, Sheffield S3 7RH, UK,\\
$^2$ Aryabhatta Research Institute of Observational Sciences
(ARIES), Nainital 263129, India
\\email:[r.j.morton, robertus]@sheffield.ac.uk}

\date{Received /Accepted}
\abstract{}{A variety of periodic phenomena have been observed in
conjunction with large solar jets. We aim to find further evidence
for {(quasi-)}periodic behaviour in solar jets and determine what
the periodic behaviour can tell us about the excitation mechanism
and formation process of the large solar jet.} {Using the
$304$~{\AA} (He-II), $171$~{\AA} (Fe IX), $193$~{\AA} (Fe XII/XXIV)
and $131$~{\AA} (Fe VIII/XXI) filters onboard the Solar Dynamic
Observatory (SDO) Atmospheric Imaging Assembly (AIA), we investigate
the intensity oscillations associated with a solar jet.}{Evidence is
provided for multiple magnetic reconnection events occurring between
a pre-twisted, closed field and open field lines. Components of the
jet are seen in multiple SDO/AIA filters covering a wide range of
temperatures, suggesting the jet can be classified as a blowout jet.
Two bright, elongated features are observed to be co-spatial with
the large jet, appearing at the jet's footpoints. Investigation of
these features reveal they are defined by multiple plasma ejections.
The ejecta display {(quasi-)}periodic behaviour {on timescales of
$50$~s} and have rise velocities of $40-150$~km\,s$^{-1}$ along the
open field lines. Due to the suggestion that the large jet is
reconnection-driven and the observed properties of the ejecta, we
further propose that these ejecta events are similar to type-II
spicules. {The bright features also display (quasi)-periodic
intensity perturbations on the timescale of $300$~s. Possible
explanations for the existence of the (quasi-)periodic perturbations
in terms of jet dynamics and the response of the transition region
are discussed.}}{}

\titlerunning{Quasi-periodicity in a blowout jet}
\authorrunning{Morton, Srivastava \& Erd\'{e}lyi }

\maketitle

\section{Introduction}
Solar jets are relatively short lived, transient, common features
often observed at the solar limb. Their presence demonstrates the
active nature of the underlying solar atmosphere suggesting
ubiquitous, fine-scale explosive magnetic events (e.g.,
\citealp{SCHetal1998}; \citealp{MOOetal2011}). The term solar jets
covers a wide range of plasma ejection events that have been
observed over the last century, including H$\alpha$ surges
(\citealp{NEW1934}), spicules (\citealp{SEC1877} or e.g.,
\citealp{BEC1972} and references within), type-II spicules
(\citealp{DEPetal2007a}), macro-spicules (\citealp{BOHetal1975}), UV
jets (\citealp{BRUBAR1983}), EUV jets (\citealp{BUDetal1998}) and
X-ray jets (\citealp{SHIetal1992}). There is some suggestion that
these events may all be closely related but no firm evidence has yet
been presented to establish this.
%#############################################################################
%#############################################################################
\begin{figure*}
\centering
\includegraphics[scale=1.0,clip=true, viewport=0.0cm 0.0cm 17.0cm 5.5cm]%
{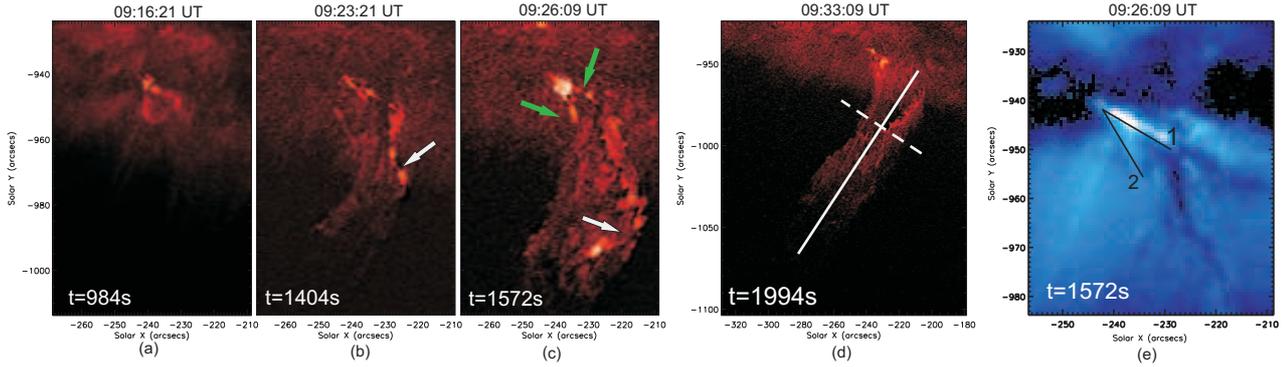} \caption{ Images from the AIA $304$~{\AA}
channel. All times in seconds are given as times after 09:00~UT on
20 January 2011. (a) A small-scale loop or arcade brightening at the
base of the solar jet. (b) The arrow shows the twisted magnetic
field highlighted by bright plasma. (c) The onset of helical motion
in the plasmas curtain. The white arrow shows the position where the
whip-like effect is seen. The green arrows point-out the two, bright
elongated features either side of the jet at its footpoints. (d) An
image of the full jet close to its maximum height. The solid white
line shows the position of the cross-cut in Fig.~\ref{fig:jet_xt}.
The dotted line is the position of the cross-cut in
Fig.~\ref{fig:jet_xt1}. (e) Shows the cross-cuts (black lines)
placed along the bright elongated features at the footpoint of the
jet in AIA $171$~{\AA}. The image also shows clearly the dark
filament thought to be an ejected magnetic loop (or arcade).
}\label{fig:jet}
\end{figure*}
%#############################################################################
%#############################################################################

Large solar jets may launch plasma hundreds of mega-meters into the
solar atmosphere, where the plasma has on occasion exhibited a clear
torsional/helical motion during the jets rise. The driving mechanism
behind these various jets is yet to be firmly established but the
height and speed of the large jets has lead to the proposition of a
mechanism involving magnetic reconnection (e.g.,
\citealp{YOKSHI1995,YOKSHI1996}). Jet models often revolve around
the emergence of magnetic flux interacting with the existing
magnetic field or the motion of existing magnetic fields
(\citealp{SHIetal1992}; \citealp{MOOetal2010}). \cite{SHIetal2007}
suggests that varying the heights of the reconnection may give rise
to the range of jets observed. Increased spatial resolution and the
ability for multi-wavelength analysis of these large jets (e.g.,
\citealp{LIUetal2011}) should provide some answers to the excitation
mechanisms and determine the main differences, if any, between the
various reported jet phenomena. Some differences between large solar
jets has been defined by \cite{MOOetal2010}, who suggest two
categories, namely, {\lq standard\rq} and {\lq blowout\rq} jets.
Standard jets fit the emerging flux-open field reconnection model
mentioned earlier, and emission is mainly seen in X-rays. On the
other hand, blowout jets begin as standard jets but the emerging
flux erupts during the formation of the standard jet. This results
in the presence of a multi-thermal jet, with chromospheric and
transition region material ejected in addition to the X-ray jet.

The importance of studying jet phenomena is highlighted by the
current interest in the so-called type-II spicules. The reported
properties of type-II spicules suggest they are fast moving
($50-150$~km\,s$^{-1}$), apparently short lived ($45-60$~s) jets
that are heated as they rise from the chromosphere, where they are
seen in Ca II filters ($T\sim0.01-0.02$~MK), to the corona, with
temperatures possibly reaching $1$~MK. The interest is due to the
suggestion that they may play a role in the heating of the quiet
Sun, acceleration of the solar wind and maintaining the mass balance
in the corona (see, e.g., \citealp{DEPetal2007,DEPetal2009};
\citealp{DEPetal2011} and \citealp{TIAetal2011}). The type-II
spicules are assumed to be excited by reconnection events in the
lower solar atmosphere and have been identified in advanced
simulations of chromospheric dynamics (\citealp{MARetal2011}).

Observations of the supposedly different jet phenomena have been
identified in a singular event. \cite{STEetal2010} report observing
bright, fast moving jet features around $20$~Mm in length in Ca II,
at the footpoints of a larger {\lq blowout\rq} solar jet seen in EUV
and soft X-rays. The authors also note the jets observed in Ca II
have rise speeds similar to type-II spicules but last for 5 minutes,
much longer than previously reported type-II events. However, the
observations offer no insight into how these Ca II jets are
generated.

%#############################################################################
%#############################################################################
\begin{figure}[t]
\centering
\includegraphics[scale=0.4,clip=true, viewport=0.0cm 0.0cm 20.0cm 14.5cm]%
{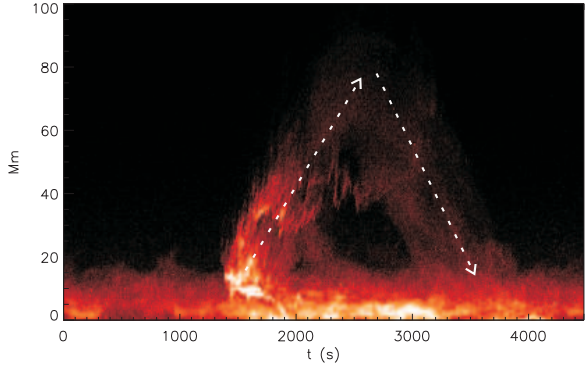} \caption{Time-distance plot from the cross-cut
along the jet axis shown in Fig.~\ref{fig:jet}d, demonstrating the
evolution of the plasma curtain. The arrows mark out the paths used
to calculate the average up-flow and down-flow speeds of the plasma
curtain. }\label{fig:jet_xt}
\end{figure}
%#############################################################################
%#############################################################################
%#############################################################################
%#############################################################################
\begin{figure}[t]
\centering
\includegraphics[scale=0.48,clip=true, viewport=0.0cm 0.0cm 18.0cm 12.0cm]%
{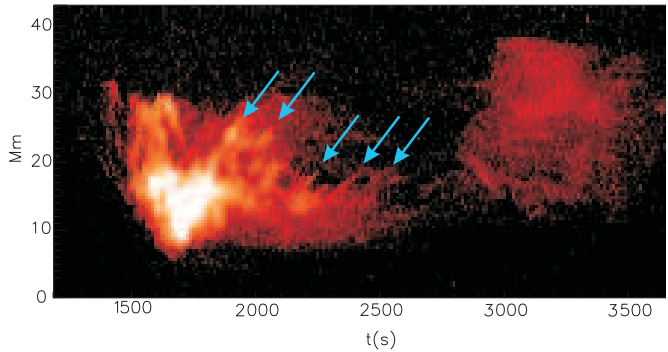}  \caption{Time-distance plot from the cross-cut
across the jet shown in Fig.~\ref{fig:jet}d. The blue arrows
highlight the diagonal tracks which are the signatures of the
torsional motion of the plasma curtain. }\label{fig:jet_xt1}
\end{figure}
%#############################################################################
%#############################################################################
Recently, the presence of waves has begun to be reported in
observations of solar jets. The study of wave and {(quasi-)}
periodic phenomena in the solar atmosphere has received increased
attention over the last two decades. This is partly due to improved
spatial and temporal resolution of both space- and ground-based
telescopes, paving the way for the new and expanding field of solar
magneto-seismology (\citealp{UCHIDA1970}; \citealp{ROBetal1984};
\citealp{ERD2006b,ERD2006}) which exploits observed wave phenomena
to determine hard to measure or otherwise unmeasurable plasma
parameters, e.g., the magnetic field strength of the individual
magnetic waveguides (see, e.g., \citealp{BANETAL2007};
\citealp{ANDetal2009}; \citealp{DEM2009}; \citealp{RUDERD2009};
\citealp{TARERD2009}; \citealp{WAN2011} for reviews of waves in the
solar atmosphere and how to exploit them for diagnostic purposes).
Advances in ground-based technology has led to the development of
high resolution and high cadence instruments that have allowed the
direct detections of, e.g., linear {torsional} Alfv\'{e}n
(\citealp{JESetal2009} ) and sausage (\citealp{MORetal2011}) waves.

Magnetic reconnection models predict that some form of {periodic}
wave phenomena is generated by a reconnection event. Two-dimensional
numerical simulations by \cite{YOKSHI1996} demonstrate the
excitation of non-linear slow and fast modes due to the
reconnection. \cite{PARetal2009} predicts the presence of kink
motions propagating along the field lines surrounding a rotating,
closed magnetic region. The kink motions are assumed to be due to a
kink instability of the magnetic region before reconnection occurs.
%#############################################################################
%#############################################################################
\begin{figure*}[!htbp]
\centering
\includegraphics[scale=0.9,clip=true, viewport=0.0cm 0.0cm 17.0cm 7.5cm]%
{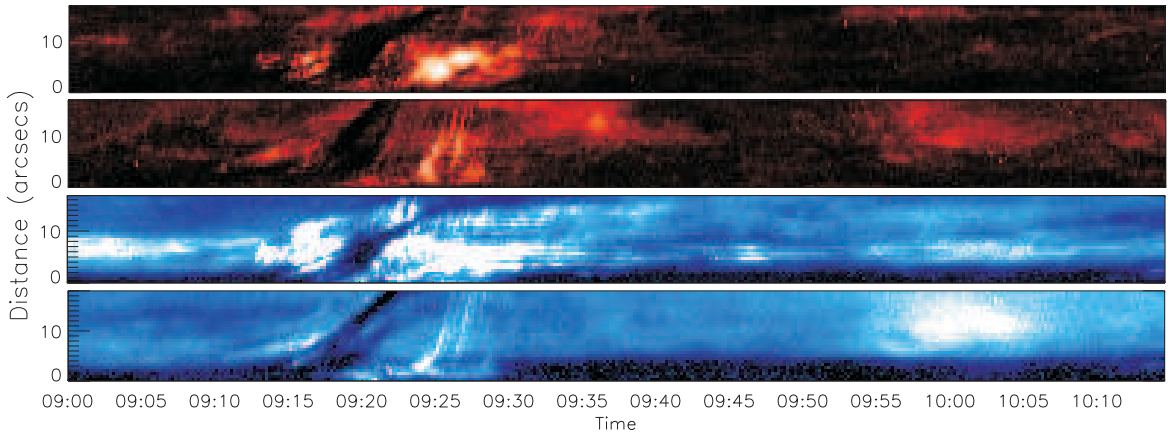} \caption{The time-distance plots for the cross-cuts
placed along the bright, elongated features at the base of the jet.
The top two panels shows the cross-cuts $No.1$ and $No.2$ obtained
for $304$~{\AA}. The bottom two panels show the cross-cuts $No.1$
and $No.2$ obtained for $171$~{\AA}.}\label{fig:slits}
\end{figure*}
%#############################################################################
%#############################################################################

The most visible, and consequently most reported, sign of waves in
the large jets is torsional/helical mode that is due to the
untwisting of the jet as it rises (e.g., \citealp{SHIMetal1996};
\citealp{PIKMAS1998}; \citealp{PATetal2008}; \citealp{LIUetal2009};
\citealp{KAMetal2010}). This feature is possibly due to the
relaxation of a twisted field line which has reconnected with an
open coronal field line (\citealp{SHIUCH1985};
\citealp{PARetal2009}). However, the helicity is also reported in
spicules and macro-spicules, e.g., \cite{ZAQERD2009};
\cite{CURTIA2011}, with spicules (type-I) generally thought to be
driven by waves (\citealp{DEPetal2004}). Transverse motions have
also been reported in both smaller jets, i.e., spicules ({see, e.g.,
\citealp{DEPetal2007}; \citealp{ZAQERD2009};
\citealp{KURetal2012}}), and large jets (\citealp{CIRetal2007};
\citealp{VAGetal2009}; \citealp{MORetal2011b}). Recent observations
and simulations in \cite{SCUetal2011} also find waves propagating
\textit{along} the transition region due to the interaction of a jet
with the transition region.

In the following work, we investigate {(quasi-)periodic} phenomena
associated with a large solar jet. Plasma with a range of
temperatures is seen to be ejected into the corona. Two, bright
elongated features at at the footpoints of the large jet are studied
and display quasi-periodic behaviour at different, distinct
timescales. The shortest timescale resolved is of the order of
$50$~s and corresponds to fast moving, bright ejecta. The measured
velocities ($40-150$~km\,s$^{-1}$), curved paths seen in
time-distance plots and lifetimes of the bright ejecta that make up
the elongated features are very similar to the previously reported
features of type-II spicules. Further, {quasi-}periodic intensity
disturbances with a longer timescale ($300$~s) are also seen in the
bright elongated features. This {quasi-periodic} behaviour is in
conjunction with a propagating transverse kink wave seen in a cool
filament ejected with the large jet (\citealp{MORetal2011b}). These
observed periodic phenomena allow us to {conjecture}, along with
observational evidence, that the large jet was excited by a low
atmosphere (possibly chromospheric) reconnection event. {We also
discuss possible mechanisms responsible for the presence of periodic
phenomena.}

\section{Observations}
The observations began at {09:00}~UT on $20$ January $2011$ and last
an hour and a quarter till {10:15~UT}, on the south east limb using
the Solar Dynamic Observatory (SDO) Atmospheric Imaging Assembly
(AIA) (\citealp{LEMetal2011}). SDO/AIA has a pixel size of
$\sim$0.6~arcsec ($435$~km) and a cadence of $\sim$12~s. The large
solar jet (Fig.~\ref{fig:jet}d) was observed clearly in the
$304$~{\AA} and $171$~{\AA} filters close to the southern polar
coronal hole. The jet event is seen partially in the higher
temperature filters of AIA, but we will only use information from
the $193$~{\AA} and $131$~{\AA} filters. The data were obtained from
the SSW cutout service and had already been corrected for
flat-field, de-spiked and co-aligned. Some aspects of the jet have
already been studied in \cite{MORetal2011b}.

The observed evolution of the jet is as follows. A region close to
the limb, approximately $11000$~km in diameter, is seen to undergo a
brightening at 09:15~UT in $171$~{\AA}. In the the same region in
$304$~{\AA}, 3-4 small patches of bright emission are simultaneously
seen, followed by the appearance of a bright loop or arcade in both
channels (Fig.~\ref{fig:jet}a). The emission is seen to increase in
the left-hand leg of the loop and then fills the loop in $\sim60$~s,
suggesting heated plasma flows along the loop.

After the loop/arcade brightening, plasma begins to rise from the
left hand side of the loop in $304$~{\AA}. The plasma rises above
the surrounding spicular material obscuring the loop from view.
Approximately $250$~s after the loop has disappeared, the plasma has
formed a collimated jet and brighter plasma starts to travel along
the edges of the jet. The bright plasma appears to highlight twisted
magnetic field lines, seen by the bright {\lq s\rq} shapes
(Fig.~\ref{fig:jet}b). As the large jet rises higher, the upper
portion of the bright twisted field lines fade and two bright,
elongated features remain at the base of the jet
(Fig.~\ref{fig:jet}c and e). These features exist while plasma is
being ejected into the atmosphere; however, once plasma ejection has
diminished they fade from view. {We note here the bright features
are an intrinsic part of the large jet.}
%#############################################################################
%#############################################################################
\begin{figure}[!htbp]
\centering
\includegraphics[scale=0.9,clip=true, viewport=0.0cm 0.0cm 10.0cm 5.5cm]%
{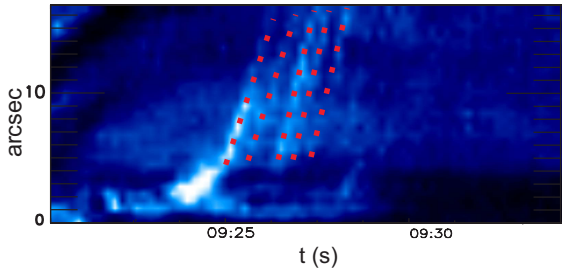} \caption{A zoom of the time-distance plot in
Fig.~\ref{fig:slits} for the cross-cut $No.2$ in $171$~{\AA}. The
paths of the multiple bright ejecta are highlighted with dotted
lines. }\label{fig:ejecta_zoom}
\end{figure}
%#############################################################################
%#############################################################################

%#############################################################################
%#############################################################################
\begin{figure}[!htbp]
\centering
\includegraphics[scale=0.7,clip=true, viewport=0.0cm 0.0cm 10.0cm 10.5cm]%
{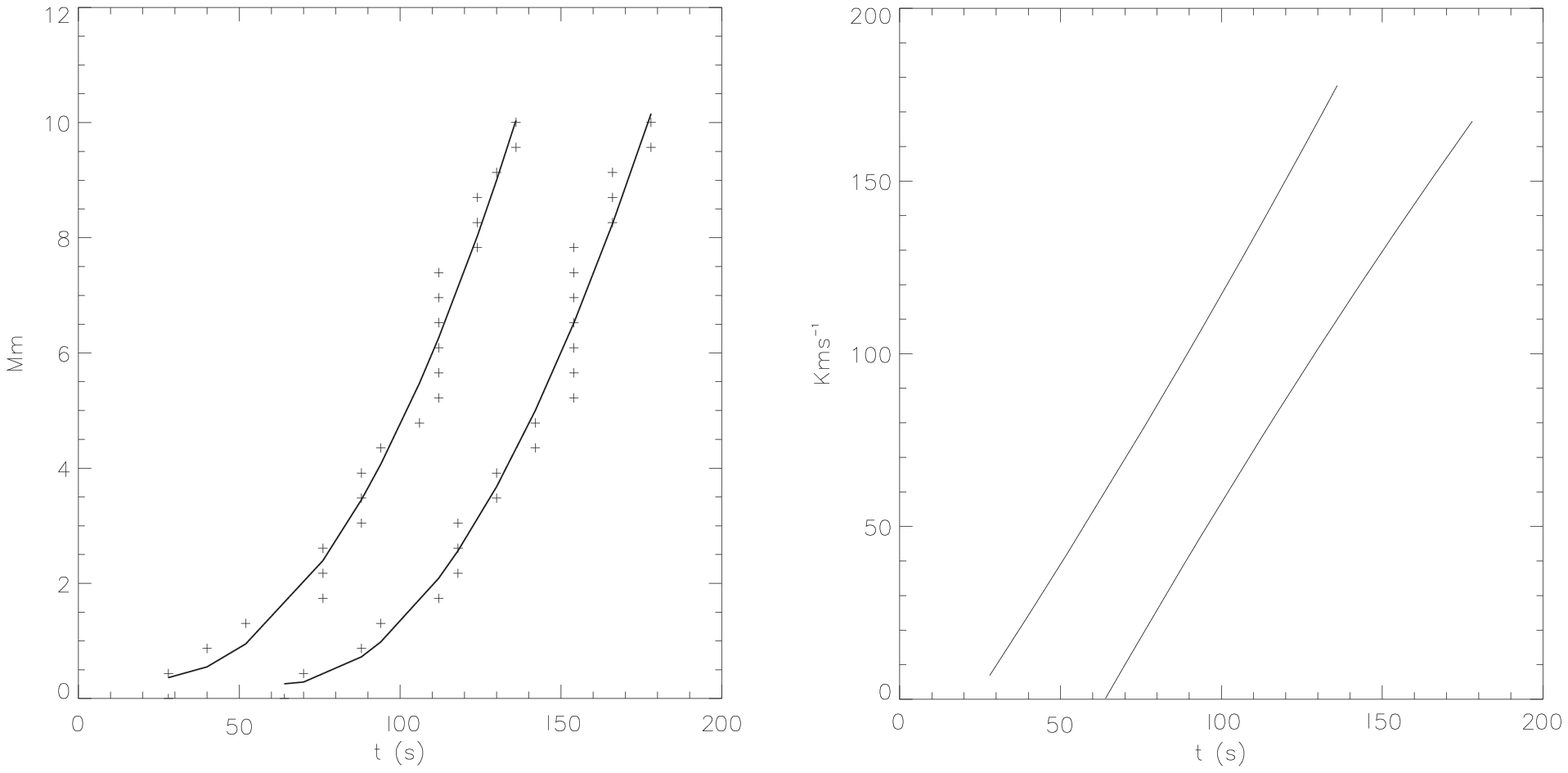}  \caption{Examples of the tracks made by the bright
ejecta seen in the time-distance plots. The crosses represent the
points of maximum intensity along the track and the solid line is a
cubic polynomial fitted to the data points. The time referred to in
this plot is unrelated to the large jet phenomena and should only be
used in relation to the properties of the ejecta, i.e. lifetimes,
velocities. }\label{fig:tracks}
\end{figure}
%#############################################################################
%#############################################################################
%#############################################################################
%#############################################################################
\begin{figure}[t]
\centering
\includegraphics[scale=0.4,clip=true, viewport=0.0cm 0.0cm 20.0cm 20.5cm]%
{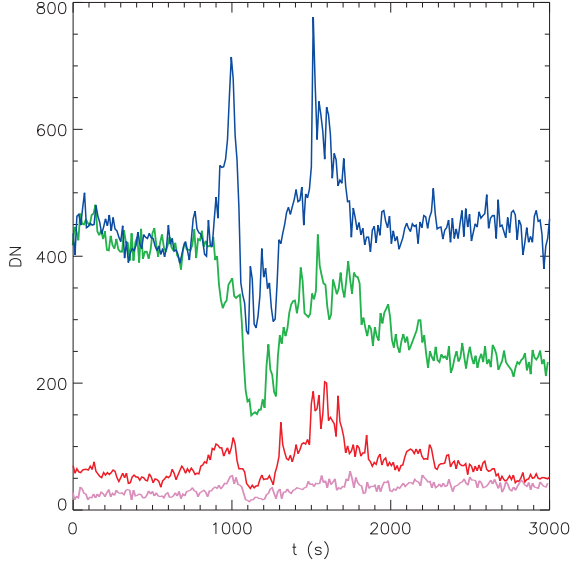} \caption{The flux as a function of time for the pixel
located at 10 arcseconds along bright feature $No.~2$ in
Fig.~\ref{fig:slits}. The red, blue, green and purple lines
corresponds to the $304$~{\AA}, $171$~{\AA}, $193$~{\AA} and
$131$~{\AA} time series, respectively. The time series are shown
from 09:00~UT to 09:48~UT. }\label{fig:flux}
\end{figure}
%#############################################################################
%#############################################################################

{The jet has a different appearance in the $304$~{\AA} channel
compared to that seen in the $171$~{\AA}, $193$~{\AA} and
$131$~{\AA}. The jet is the most visible in $304$~{\AA}, where a
significant amount of plasma is seen to be launched into the solar
atmosphere to a height of $\sim100$~Mm. This material makes up what
is referred to as the plasma curtain. The ejected plasma flows
outwards, at an angle to the normal from the surface (from left to
right in Fig.~\ref{fig:jet}c). At a later time, a portion of the
ejected material is also seen to return to the surface. A cross-cut
along the axis of the jet is made and a time-distance plot of the
jet rise is obtained (Fig.~\ref{fig:jet_xt}). Bright, outward
flowing plasma can be seen to be initially ejected in bursts at
$\sim1400$~s (i.e. $1400$~s after 09:00~UT). An average value for
the outwards velocity of the plasma during the rise phase is
$72$~km\,s$^{-1}$. The plasma curtain cools and begins to fall, with
an average speed of $82$~~km\,s$^{-1}$, giving the observed
parabolic profile.}

{At a height of $\sim18$~Mm above the jet base, a whip-like motion
of the the jet occurs sending the out-flowing plasma in the opposite
direction (right to left), giving the plasma curtain a helical
motion (Fig.~\ref{fig:jet}c). Another cross-cut (see,
Fig.~\ref{fig:jet}d for cross-cut position) is placed perpendicular
to the jet axis at around $33$~Mm above the jet footpoint. The
corresponding time-distance plot is shown in Fig.~\ref{fig:jet_xt1}.
The onset of helical motion in the plasma curtain can be seen at
$1700$~s, diagonal tracks in the time-distance plot made by the
plasma are the result. Some of these tracks are highlighted by the
arrows in the plot.}

{The plasma curtain is not readily visible in the $171$~{\AA},
$193$~{\AA} and $131$~{\AA} channels but faint, co-spatial
up-flowing emission is observed. It can be seen in
Fig.~\ref{fig:jet} that the plasma curtain has similar emission to
the surrounding spicular material, with a number of regions
displaying enhanced emission. This would suggest the temperature of
the majority of the plasma curtain is at chromospheric/transition
region temperatures; however, the observed enhanced emission would
mean the plasma is likely to be multi-thermal with a wide range of
temperatures.} The bright elongated features are still present {in
the $171$~{\AA}, $193$~{\AA} and $131$~{\AA} channels} along with a
dark filament. In $193$~{\AA} and $131$~{\AA} only the bright,
elongated feature (labeled $1$) closest to the solar surface is
clearly visible, the other feature ($2$) is much fainter in
comparison.

The properties of the dark filament are explored in detail in
\cite{MORetal2011b}. The filament is seen to rise from the jet
excitation site and magneto-seismology suggests it has a temperature
of $2-3\times10^4$~K. {There is evidence that the filament is
partially seen in the $304$~{\AA} channel, again as an absorbing
feature. However, the plasma curtain forms a {\lq sheath\rq} around
the filament and partially obscures it from sight.} This suggests
the filament is the remnant of a chromospheric magnetic loop (or
arcade) that has been ejected.

The bright loop/arcade and the twisted field lines suggest a
scenario where twisted, emerging flux has reconnected with open
field lines and heated plasma flows along the open lines. This
scenario has been suggested previously for the formation of large
jets (\citealp{SHIetal1992}) and preliminary studies of this
scenario are being carried out with numerical simulations (e.g.,
\citealp{PARetal2009}). However, we also see cool chromospheric
material ($T\sim0.03$~MK) in the form of the dark filament and upper
chromospheric/transition region material in the plasma curtain. This
would indicate that we observe an example of a blowout jet, as
suggested by \cite{MOOetal2010}. We note also that the spire,
typically associated with plasma heated by reconnection, is absent
from the event.

%#############################################################################
%#############################################################################
\begin{figure*}[t]
\centering
\includegraphics[scale=0.7,clip=true, viewport=0.0cm 0.0cm 20.0cm 12.5cm]%
{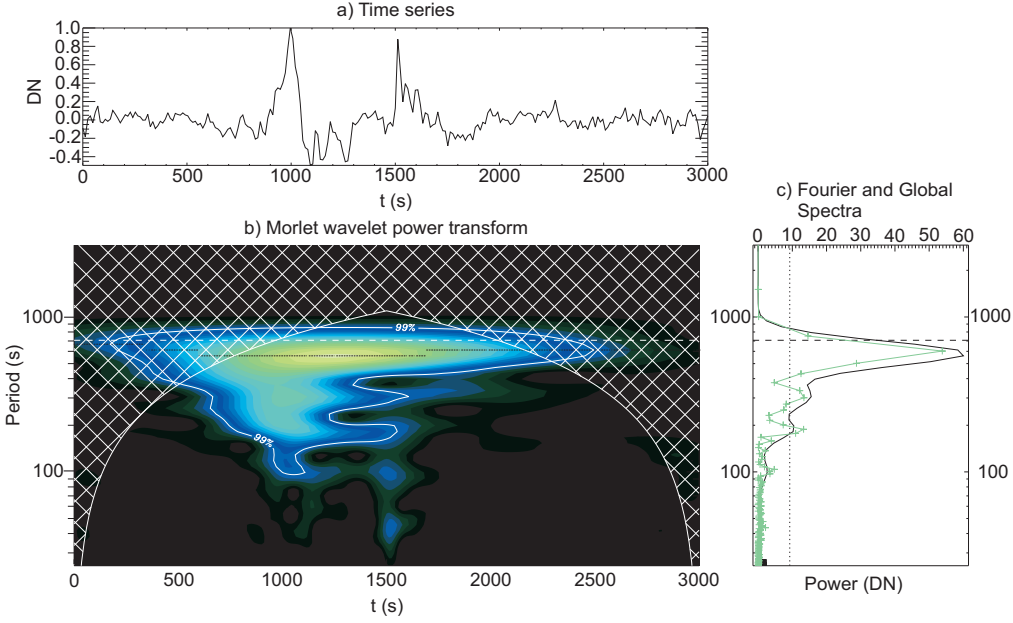} \caption{(a) The time series for the normalised
intensity in $171$~{\AA} located at 10 arcseconds in cross-cut
$No.2$ shown in Fig.~\ref{fig:slits}. The time is measured in
seconds from 09:00~UT on 20 January 2011. (b) The wavelet plot for
the time series with regions of significant power ($>99$\%)
contoured with the white line. The coloured regions show power above
$95\%$. (c) The Fourier and global powers spectra for the analysed
time series.}\label{fig:wave}
\end{figure*}
%#############################################################################
%#############################################################################
%#############################################################################
%#############################################################################
\begin{figure*}[t]
\centering
\includegraphics[scale=0.7,clip=true, viewport=0.0cm 0.0cm 20.0cm 12.5cm]%
{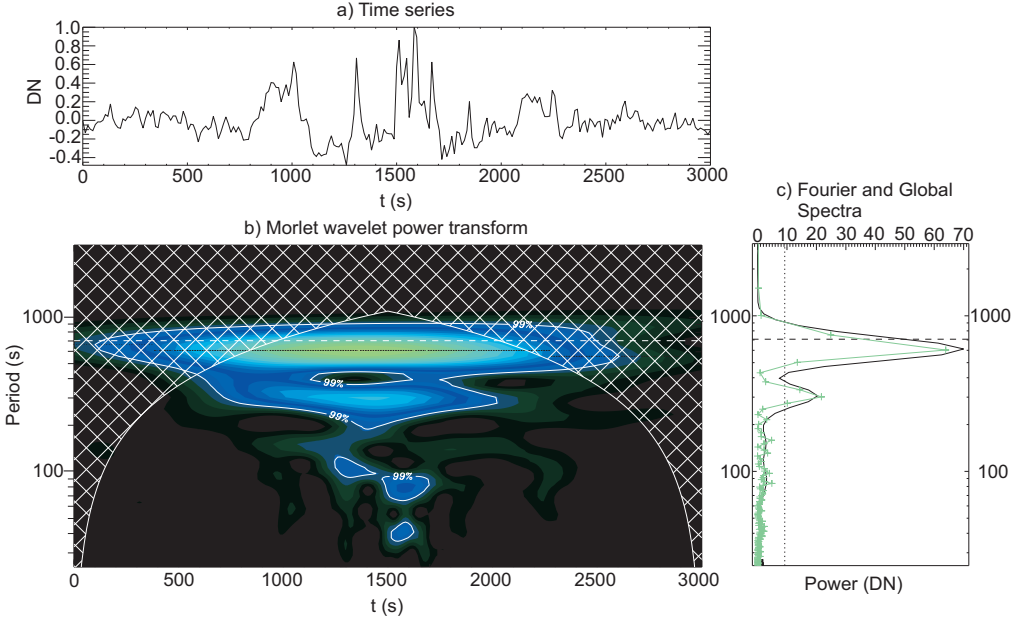}  \caption{Similar to Fig.~\ref{fig:wave} expect
the analysed time series is from the $304$~{\AA}
cross-cut.}\label{fig:wave2}
\end{figure*}
%#############################################################################
%#############################################################################

\section{Periodic behaviour}
Now that a description of the large jet has been given, let us
proceed to examine the blowout jet event in detail. In particular,
we are interested in the bright, elongated features seen at the
footpoints of the jet. {These features are intrinsically related to
the jet formation process and the mass flow into the corona.}
Further, we search for periodic phenomena in the intensity time
series of plasma associated with the bright elongated features. To
our knowledge, the investigation of intensity perturbations
associated with large jets has so far been neglected. {We aim to
resolve this and discuss their presence in terms of the formation
process of the bright features and the large jet.}

\subsection{Short Timescales}
To begin with, we place cuts ($1$-pixel wide, $30$ pixels in length)
along the bright, elongated features seen at the foot-points of the
jet (Fig.~\ref{fig:jet}e). The bright features are around $12$~Mm in
length and separated by $20$ degrees, with all the ejected plasma
that comprises the large jet appearing from between these two
features. The features show increased emission (compared to the
surrounding plasma) for around $400$~s. In Fig.~\ref{fig:slits} we
show the time-distance plots obtained from the cross-cuts placed
along the bright features in the $304$~{\AA} and $171$~{\AA}
channels. In both channels and both cuts we see increased emission
around 09:15~UT indicating the onset of the jet. This is followed by
the rise of {fainter} plasma, then the rise of further plasma with
increased emission.

In the cross-cuts placed along the bright feature labeled $2$
(Fig.~\ref{fig:jet}e), it can be seen that there are multiple,
bright ejections starting at 09:23~UT. The exact number of ejections
is hard to distinguish but it appears to be at least {five}
occurring within a $200-300$~s time frame. A zoom of these ejecta as
seen in the $171$~{\AA} cross-cut are shown in
Fig.~\ref{fig:ejecta_zoom} with dotted lines highlighting the paths
of the brightest ejecta. It is evident from the time-distance plots
that the features are fast moving but the ejections do not have a
constant speed. The first ejection is the brightest and it can be
seen that, close to the base of the cross-cut, the path is curved,
suggesting acceleration. An example of two ejecta paths are shown in
Fig.~\ref{fig:tracks}. The tracks where obtained by determining the
pixels with the greatest value of intensity in the time-distance
plots and fitting a cubic polynomial to the obtained points. The
ejection has an initial speed of $40$~km\,s$^{-1}$ and accelerates
to $150$~km\,s$^{-1}$. The bright material fades as it reaches the
end of the cross-cut. The lifetime of the individual ejecta in both
$304$~{\AA} and $171$~{\AA} filters is between $30-120$~s. {This
behaviour is not visible in the cross-cut for the bright feature
No.~1 (Fig.~\ref{fig:slits}), however, the provided online movies
shows many similar ejections from the bright feature.}

{Now, let us examine the flux time series and search for the
signature of these fast moving ejections.} We select the pixels
located at $10$~arcseconds (mid way) along the cross-cut $No.~2$ in
$304$~{\AA}, $171$~{\AA}, $193$~{\AA} and $131$~{\AA}. The time
series for the intensity between 09:00~UT and 09:48~UT of these
pixels are shown in Fig.~\ref{fig:flux}. {The plotted flux profile
is typical of the other pixels in the cross-cut}. At $1500$~s we see
five large peaks in intensity with a quasi-periodicity on $\sim50$~s
{timescales} in {three of the four selected filters}. To investigate
further, we de-trended the time series for $171$~{\AA} and
$304$~{\AA} and subject it to a wavelet transform. The results are
plotted in Figs.~\ref{fig:wave} and \ref{fig:wave2}. Significant
power (contoured at $99\%$) is found in the global spectra at
timescales of $\sim200$~s and $\sim300$~s in $171$~{\AA}
(Fig.~\ref{fig:wave}) and at $\sim300$~s in $304$~{\AA}
(Fig.~\ref{fig:wave2}). The wavelet transforms
(Figs.~\ref{fig:wave}b, \ref{fig:wave2}b) also show the presence of
power at $50-100$~s {timescales} during the time of the ejecta.
However, the power is contoured above $99\%$ significance only in
the wavelet of the $304$~{\AA}. The wavelet for $171$~{\AA} shows
power at $95\%$ for $50-100$~s timescales.

\begin{figure*}[t]
\centering
\includegraphics[scale=0.8,clip=true, viewport=0.0cm 0.0cm 20.0cm 12.5cm]%
{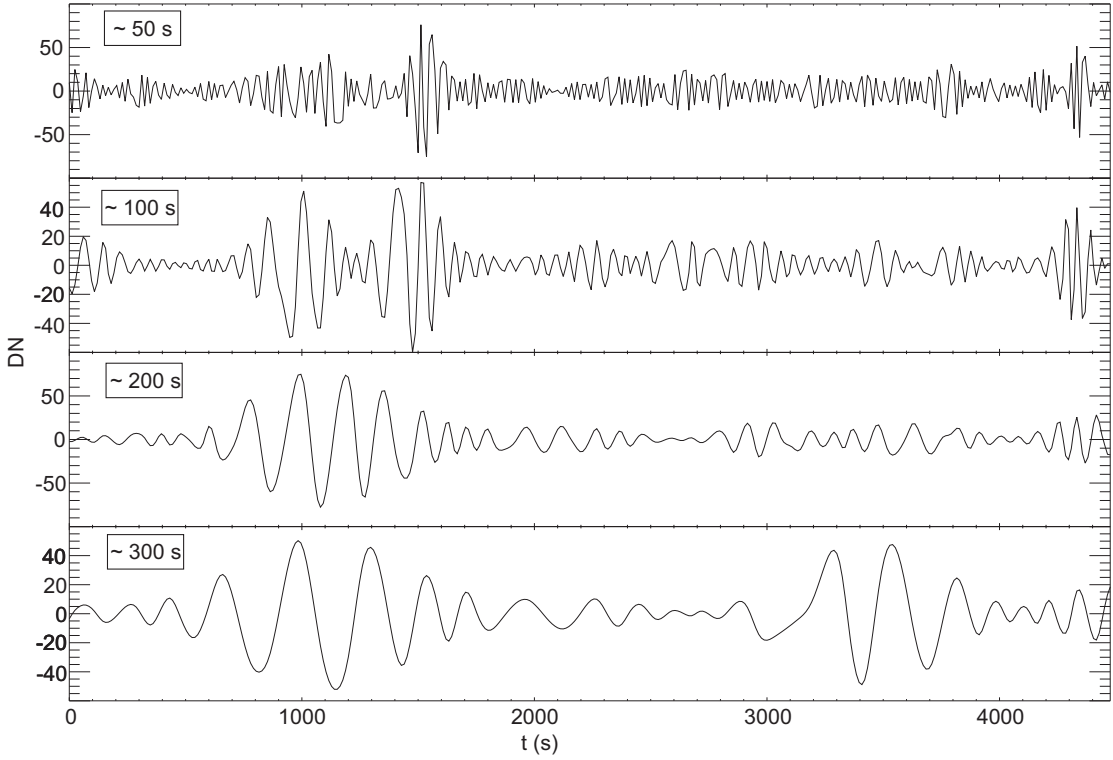}  \caption{The first four Intrinsic modes functions
obtained using Empirical mode decomposition. From top to bottom, the
panels correspond to characteristic time scales of $50$~s, $100$~s,
$200$~s and $300$~s.}\label{fig:emd}
\end{figure*}

\begin{figure}[t]
\centering
\includegraphics[scale=0.65,clip=true, viewport=0.0cm 0.0cm 14.0cm 8.5cm]%
{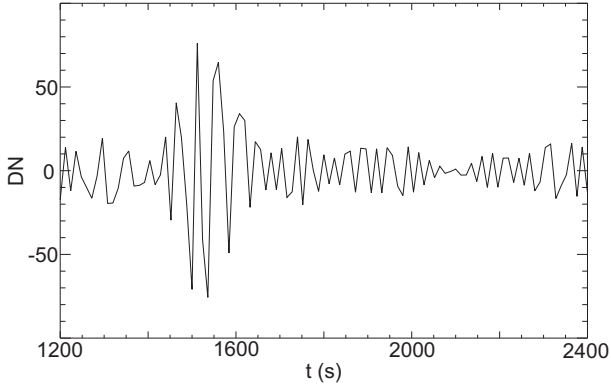}  \caption{Zoom of the first IMF shown in
Fig.~\ref{fig:emd} to highlight the {quasi-periodic phenomena with a
timescale of} $50$~s in the bright emission.}\label{fig:emd_1}
\end{figure}

To extract the feature we are interested in, we now turn to
Empirical Mode Decomposition (EMD) (for details see, e.g.,
\citealp{HUAetal1998}; \citealp{TERetal2004}). The first four
intrinsic mode functions (IMFs) derived with the EMD for $171$~{\AA}
are shown in Fig.~\ref{fig:emd}. They correspond to timescales of
$50$~s, $100$~s, $200$~s and $300$~s identified in the wavelet
transforms. Let us concentrate on the first IMF. At $1500$~s we see
the large peaks in intensity, also observed in the flux time series
(Fig.~\ref{fig:flux}). This is the signal of the fast moving,
ejected material and can be seen to be periodic (or at least
quasi-periodic) with a decaying amplitude. A zoom of the first IMF
between $1200-2400$~s is shown in Fig.~\ref{fig:emd_1}. The
magnitude of the amplitude of the signal is $18\%$ of the median
background signal in $171$~{\AA} and $41\%$ of the median background
signal in $304$~{\AA}.

{To observe this behaviour as a function of distance along the
cross-cut,} a Fast Fourier Transform (FFT) for each pixel in the
cross-cut is taken and a Gaussian filter centered on $50$~s with a
width $f/10$ is applied and the inverse FFT is taken. The results
are shown in the top panels of Figs.~\ref{fig:filt_171} and
\ref{fig:filt_304}. This plot provides a somewhat better
visualisation of the high speed, up-flowing material, where the
curved tracks show a visibly coherent signal along the entire
cross-cut between $1400$~s and $1800$~s. Although we cannot see the
individual ejections in Fig.~\ref{fig:slits} for the bright feature
$1$, the EMD and the FFT technique do reveal that there are similar,
fast moving, up-flowing features with a {timescale} of $50$~s.

%#############################################################################
%#############################################################################
%#############################################################################
%#############################################################################
\subsection{Longer Timescales}
\begin{figure*}[!htbp]
\centering
\includegraphics[scale=0.9,clip=true, viewport=0.0cm 0.0cm 20.0cm 6.cm]%
{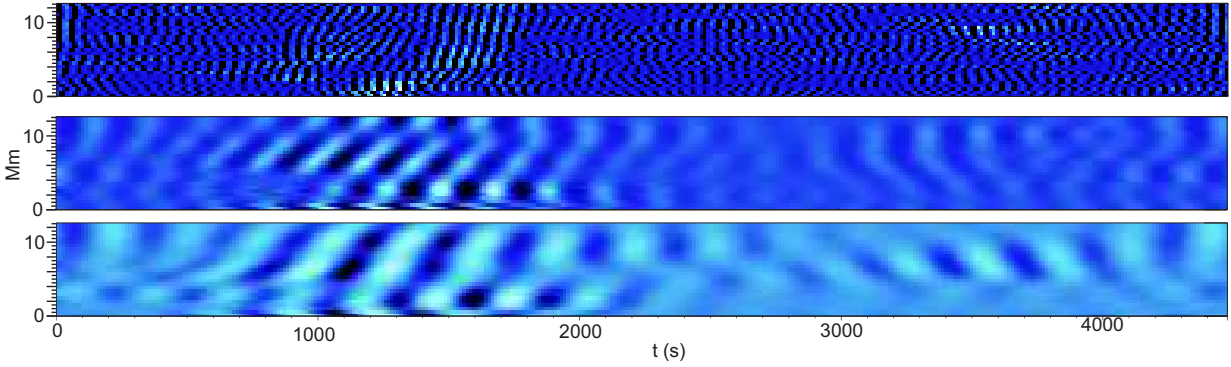}  \caption{Frequency filtered time series
obtained from the $171$~{\AA} cross-cut $No.2$ shown in
Fig.~\ref{fig:slits}. The top, middle and bottom rows are filtered
at {timescales of} $50$~s, $200$~s and $300$~s,
respectively.}\label{fig:filt_171}
\end{figure*}

\begin{figure*}[!htbp]
\centering
\includegraphics[scale=0.9,clip=true, viewport=0.0cm 0.0cm 20.0cm 4.cm]%
{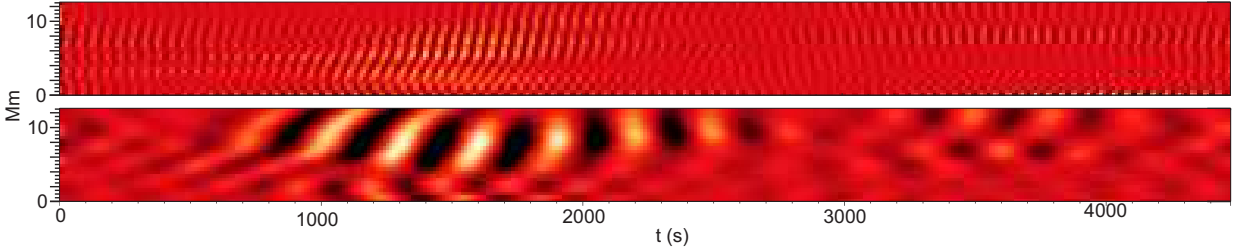}  \caption{Frequency filtered time series
obtained from the $304$~{\AA} cross-cut $No.2$ shown in
Fig.~\ref{fig:slits}. The top and bottom rows are filtered at
{timescales of} $50$~s and $300$~s,
respectively.}\label{fig:filt_304}
\end{figure*}

Now, let us investigate the nature of the phenomena with timescales
$>50$~s which had significant power in the wavelets
(Fig.~\ref{fig:wave} and \ref{fig:wave2}). We ignore the timescale
peaked at $600$~s, as this begins well before and ends after the
large jet. Further, the Fourier power shows contributions from a
wide range of timescales ($400-900$~s). Instead, we first
concentrate on the phenomena with a {timescale} of $300$~s, which is
found in the wavelets of the intensity time series from both
$304$~{\AA} and $171$~{\AA}. We again apply a Gaussian filter (as
discussed before) centered on $300$~s with a width $f/10$. The
results are plotted in Fig.~\ref{fig:filt_171} (third panel) for
$171$~{\AA} and Fig.~\ref{fig:filt_304} (second panel) for
$304$~{\AA}. A coherent, {quasi-periodic} signal (after $1000$~s)
that propagates along the cross-cut can be seen in both channels.
The measured speed of this quasi-periodic feature along the
cross-cut is $35\pm5$~km\,s$^{-1}$.

\begin{figure}[!htbp]
\centering
\includegraphics[scale=0.8,clip=true, viewport=0.0cm 0.0cm 11.0cm 8.cm]%
{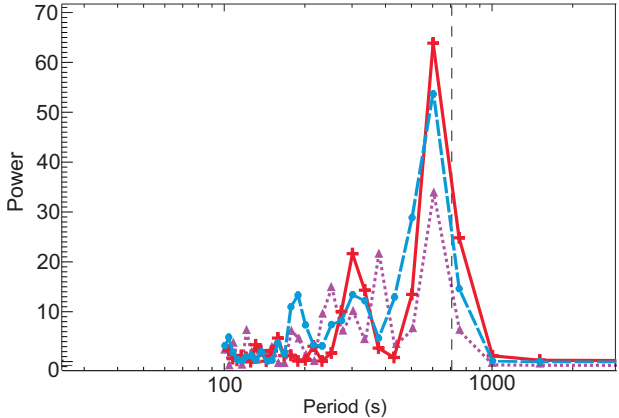}  \caption{Comparison of Fourier power derived
from wavelet routine. The power series are derived from the
intensity time series for $304$~{\AA}, $171$~{\AA} and $131$~{\AA}
shown in red (solid line and crosses), blue (dash and circles) and
purple (dot and triangles), respectively.}\label{fig:amp_comp}
\end{figure}

We have also applied the Gaussian filtering process to the entire
data cube that makes up the jet observation, for both $171$~{\AA}
and $304$~{\AA}. The movies for this are supplied online
($http://swat.group.shef.ac.uk/current\_work.html$). {In the movie
of the filtered $304$~{\AA} channel, the spicular material
surrounding the jet is also seen as it has typical timescales of
$300$~s.} {The quasi-periodic phenomena is found throughout the
large jet and is dominant in the region defined by the plasma
curtain (as seen in the $304$~{\AA} images). The quasi-periodic
behaviour has relatively large amplitudes at the footpoint of the
jet but the amplitude decreases with height (also seen in the
cross-cuts, Figs.~\ref{fig:filt_171} and \ref{fig:filt_304}).
Comparing the Fourier power spectra derived from the wavelets
(Fig.~\ref{fig:amp_comp}), we note that the power of the
quasi-periodic phenomena is stronger in the $304$~{\AA} channel with
decreasing power in the $171$~{\AA} and $131$~{\AA} channels. This
feature is also seen by comparing the movie of the filtered
$171$~{\AA} images to the filtered $304$~{\AA} images.}

%#############################################################################
%#############################################################################
%#############################################################################
%#############################################################################

\vspace{0.1cm} Finally, we discuss the intensity perturbation with a
{timescale of} $200$~s, observed only in the $171$~{\AA} wavelet
transform. We again apply a Gaussian filter about $200$~s for each
pixel in cross-cut $2$ (results shown in middle panel of
Fig.~\ref{fig:filt_171}). A coherent signal is seen to propagate
along the slit, but only over a distance of $\sim6-7$~Mm. The time
at which the intensity perturbation begins is at the same time as
the dark filament passes through the cross-cut. The dark features
and intensity perturbations rise velocities along the cross-cut are
$\sim23$~km\,s$^{-1}$ (the actual rise velocity of the filament is
$\sim60$~km\,s$^{-1}$, see \citealp{MORetal2011b}). Hence, it would
suggest that the passage of the filament is the cause of the
intensity perturbation. This is confirmed as we find a velocity of
$\sim16$~km\,s$^{-1}$ for the filament and the intensity
perturbation with a {timescale} of $200$~s as seen in cross-cut $1$.

The cross-cuts along the bright features are located at an angle to
the path that the dark filament makes through the atmosphere, hence
we obtain rise speeds a fraction of that reported in
\cite{MORetal2011b}. If the filament is the driver for the
{quasi-periodic perturbation}, then the propagation speed for the
intensity perturbation could also be as high as
$\sim60-70$~km\,s$^{-1}$. Placing a cross-cut along the filaments
axis, almost perpendicular to the solar surface (the orientation of
the filament can be seen in Fig.~\ref{fig:jet}e), we do indeed
obtain velocities of the order $60$~km\,s$^{-1}$ associated with the
{intensity perturbation}. {The role of the filament in exciting the
quasi-periodic phenomena is also seen in the movie of the
$171$~{\AA} images filtered at $200$~s. The amplitude and height of
the quasi-periodic phenomena can be observed to increase as the
filament rises.}

\section{Discussion}\label{sec:disc}
{The nature of the periodic signals seen in the wavelet and EMD of
the intensity are not straightforward. Possible options for the
intensity perturbations may be due to MHD waves, quasi-periodic
up-flows or dominant time-scales related to the jet event. We now
discuss the possible interpretation of each observed event. We note
there is currently some debate as to the exact differences between
the observational signatures of quasi-periodic up-flows and slow
modes. Such a discussion is outside the scope of this investigation,
so we refer an interested reader to the following pieces of work:
\cite{DEPMCI2010}; \cite{VERetal2010}; \cite{WANetal2011}.}

High-speed up-flows in large solar jets have been reported
previously by, e.g., \cite{LIUetal2009} and \cite{STEetal2010},
although no periodicity was reported in the up-flow. The properties
of the high-speed up-flows that are observed here, also bring to
mind recent results of investigations into type-II spicules (e.g.
\citealp{DEPetal2007a}). It is thought that reconnection drives the
type-II spicules, which are characterised by the ejection of high
velocity up-flows ($\sim40-120$~km\,s$^{-1}$) of hot material into
the corona (\citealp{DEPetal2011}). Previously reported type-II
spicules are seen to originate in the chromosphere with temperatures
of $\lesssim0.1$~MK and velocities of $\sim40-120$~km\,s$^{-1}$,
before appearing in coronal lines with temperatures of $0.2-1$~MK.
The presence of type-II spicule-like events in large jets are also
discussed in \citealp{STEetal2010}. As pointed out by the authors,
the type-II spicule-like features are part of the macroscopic jet so
the situation is somewhat more complicated than individual type-II
spicules.

Such a statement is highlighted with the following differences
between the spicule-like events we see here and previously reported
type-II spicules. The bright, fast moving features we observe here
have a {quasi-periodic} rate of recurrence of the order of $50$~s,
while previous reports indicate periods of $5-12$ minutes for
type-II spicules (\citealp{DEPMCI2010}). A further difference is
that the {quasi-periodic} behaviour observed here displays apparent
damping and only lasts for five periods, whereas no such features
have been reported with type-II spicules previously, to our
knowledge. However, these properties probably indicate a difference
between the drivers of the events.

How can we explain such a phenomena in terms of the suggested
mechanisms of jet formation? One option is that multiple
reconnection events between emerging, closed field lines and open
field lines take place. This would cause the numerous ejections that
are being guided upwards by the open field lines. The data also
suggests that the reconnection event would probably have to be
periodic or quasi-periodic. The possibility for periodic
reconnection has been demonstrated in 2-d models if the reconnection
has been driven by a fast magneto-acoustic wave
(\citealp{MCLetal2010}).

Another possibility is that only one reconnection event has occurred
and we are seeing the response of the transition region to a single
velocity pulse. The rebound-shock model suggests that a velocity
pulse can cause the transition region to generate a damped,
(quasi-)periodic response (\citealp{HOLetal1982};
\citealp{MURZAQ2010}). However, the bright, fast moving emission can
be seen up to heights of at least 18 arcseconds ($12$~Mm).
Simulations of spicule formation via waves show that wave driven
spicule phenomena cannot reach such heights, e.g., the simulation of
small chromospheric jets such as spicules (\citealp{DEPetal2004}),
fibrils (\citealp{HANetal2006}), mottles (\citealp{HEGetal2007})
typically have heights $<5$~Mm. {Typical rise velocities for
wave-driven spicules are also lower ($10-40$~km\,s$^{-1}$) than
those measured here.} Although, typical amplitudes and velocities
for these simulations are based on driving by photospheric
granulation, so the results may not be directly applicable here if
reconnection driven outflows or non-linear waves are the driver.
{Recent simulations of a pulse driven jet (\citealp{SRIMUR2011})
demonstrate that an initial large amplitude pulse can lead to the
ejection of material to heights of tens of Mm with rise velocities
of the order of $100$~km\,s$^{-1}$.}

{Enhanced emission at the jets footpoints is seen to exist for
longer than $200-250$~s time period associated with the initial,
fast moving ejections (although the greatest emission is observed
with these ejecta). The enhanced emission also displays
quasi-periodic intensity perturbations with the time-scale of
$300$~s. This may suggest that two separate mechanisms are
responsible for its appearance. Similar arguments to those just
given about the origins of the periodicity can be applied.}

{If we assume reconnection is responsible, than we have to explain
the presence of the quasi-periodicity. The observed timescale is
similar to that previously reported for a range of wave phenomena
present throughout the solar atmosphere see, e.g.
\citealp{BANETAL2007}. It has been demonstrated that the waves in
the lower solar atmosphere can drive reconnection to produce small
scale jet events with the same periodicity as the exciting wave
(\citealp{HEGetal2009}). On the other hand, waves generated by
reconnection (e.g. \citealp{YOKSHI1996}) could also display these
timescales.}

An alternative to reconnection is, again, the rebound shock process.
From the results of previous simulations of the rebound shock
process, it would appear more feasible that they could generate the
observed behaviour at $\sim300$~s timescales rather than the
$\sim50$~s timescale. For example, the observed up-flow speed of
$35$~km\,s$^{-1}$ is compatible with rebound-shock spicule
simulations (e.g., \citealp{MURZAQ2010}).

{The movies of the filtered time series also show that the
quasi-periodic behaviour is also present throughout the large jet.
The amplitude of the perturbations appears to decrease significantly
with height (e.g., compare amplitudes at the footpoints to the
amplitudes at $<30$~Mm above the surface). The regions which display
the (quasi-)periodic behaviour are co-spatial with the plasma
curtain. This hints that whatever is generating the observed
timescales and the bright elongated features also contributes to the
formation of the plasma curtain. Although, further investigation is
needed before a firm statement can be made on this.}
%#############################################################################
%#############################################################################
%#############################################################################
%#############################################################################

%#############################################################################
%#############################################################################
%#############################################################################
%#############################################################################

There is also the presence of an intensity perturbation with a
period of $200$~s in the $171$~{\AA} time series. The
quasi-periodicity appears to be driven by the passage of the cool
filament. To explain the {quasi-periodic} nature of the signal, we
suggest the passage of the filament through the transition region
could result in the initiation of the rebound-shock process. The
filament fills the role of a density/velocity pulse, as used in the
simulations, and initiates {quasi-periodic} behaviour as it passes
through the transition region. The amplitude of the signal for the
perturbation is larger close the jet footpoints but appears to decay
with height and does not propagate high into the atmosphere (see
online movie). The amplitude is also seen to decay with time in the
EMD results (Fig.~\ref{fig:emd}, third panel). Similar behaviour is
also found in the rebound-shock simulations. If this is indeed the
scenario, then the observed quasi-periodic behaviour, i.e., the
intensity perturbation with a period of $200$~s, would not then tell
us anything about the initiation process of the large jet.

%#############################################################################
%#############################################################################
%#############################################################################
%#############################################################################

\section{Summary and Conclusions}
We study a large solar jet event that occurs close to the southern
polar coronal hole. Due to a number of observational signatures such
as a loop/arcade brightening, an untwisting field and the final
height of the large jet we conclude that the jet is probably
initiated by a magnetic reconnection event between a twisted, closed
field and an open magnetic field. The presence of plasma with a wide
range of temperatures suggests this event falls into the blowout jet
scenario (\citealp{MOOetal2010}), although no high temperature
(X-ray) component is observed. To investigate the process of jet
formation we take a novel approach and look for periodic behaviour
in intensity associated with enhanced emission at the large jet's
footpoints.

{For part of the enhanced emissions lifetime}, bright elongated
features composed of multiple bright, fast moving
($40-150$~km\,s$^{-1}$) plasma ejections are observed. The ejections
have properties that are similar to those reported for type-II
spicules. Interpretation of the observed events as type-II
spicule-like features is not incompatible with the idea that a
reconnection-based mechanism is responsible for generating the large
solar jets. The events appear to have a timescale of $50$~s,
suggesting that (quasi-)periodic reconnection may be occurring.

\vspace{0.1cm}On the other hand, there exists quasi-periodic
intensity perturbations with a timescale of $300$~s in the enhanced
emission. The oscillation had significant power in the $304$~{\AA}
and $171$~{\AA} channels and was found to propagate along the slits
at $35$~km\,s$^{-1}$. The appearance of two separate time-scales,
i.e., 50~s and 300~s, could suggest that more than one mechanism is
responsible for determining the jet dynamics.

\vspace{0.1cm} In our view, the presence of the multiple bright,
fast moving plasma ejections here, and in previous works
(\citealp{STEetal2010}), suggests that type-II spicule-like events
could play an important role in determining the appearance of the
jets. Type-II events are already thought to be important for
maintaining the mass balance in the corona and could be important
for the distribution of the UV/EUV temperature material, that makes
up the larger jet, into the corona. The extent of the role played by
the response of the transition region to waves and flows impinging
on it also needs further detailed study with regards to jet
formation.

\begin{acknowledgements}
The authors thank the referee for providing insightful comments that
improved the manuscript and A. Engell, D. Jess, and P. Grigis for a
number of useful discussions. RE acknowledges M. K\'eray for patient
encouragement. The authors are also grateful to NSF, Hungary (OTKA,
Ref. No. K83133) and the Science and Technology Facilities Council
(STFC), UK for the support they received. AKS thanks SP$^2$RC for
the financial support received for the visits and collaborative
research. AKS also acknowledges the encouragement and support of
Shobhna Srivastava. The data is provided courtesy of NASA/SDO and
the AIA, EVE, and HMI science teams.
\end{acknowledgements}

\bibliographystyle{aa}

\begin{thebibliography}{61}
\expandafter\ifx\csname
natexlab\endcsname\relax\def\natexlab#1{#1}\fi

\bibitem[{{Andries} {et~al.}(2009){Andries}, {van Doorsselaere}, {Roberts},
  {Verth}, {Verwichte}, \& {Erd{\'e}lyi}}]{ANDetal2009}
{Andries}, J., {van Doorsselaere}, T., {Roberts}, B., {et~al.} 2009,
Space
  Science Reviews, 149, 3

\bibitem[{{Banerjee} {et~al.}(2007){Banerjee}, {Erd{\'e}lyi}, {Oliver}, \&
  {O'Shea}}]{BANETAL2007}
{Banerjee}, D., {Erd{\'e}lyi}, R., {Oliver}, R., \& {O'Shea}, E.
2007,
  \solphys, 246, 3

\bibitem[{{Beckers}(1972)}]{BEC1972}
{Beckers}, J.~M. 1972, \araa, 10, 73

\bibitem[{{Bohlin} {et~al.}(1975){Bohlin}, {Vogel}, {Purcell}, {Sheeley},
  {Tousey}, \& {Vanhoosier}}]{BOHetal1975}
{Bohlin}, J.~D., {Vogel}, S.~N., {Purcell}, J.~D., {et~al.} 1975,
\apjl, 197,
  L133

\bibitem[{{Brueckner} \& {Bartoe}(1983)}]{BRUBAR1983}
{Brueckner}, G.~E. \& {Bartoe}, J. 1983, \apj, 272, 329

\bibitem[{{Budnik} {et~al.}(1998){Budnik}, {Schroeder}, {Wilhelm}, \&
  {Glassmeier}}]{BUDetal1998}
{Budnik}, F., {Schroeder}, K., {Wilhelm}, K., \& {Glassmeier}, K.
1998, \aap,
  334, L77

\bibitem[{{Cirtain} {et~al.}(2007){Cirtain}, {Golub}, {Lundquist}, {van
  Ballegooijen}, {Savcheva}, {Shimojo}, {DeLuca}, {Tsuneta}, {Sakao}, {Reeves},
  {Weber}, {Kano}, {Narukage}, \& {Shibasaki}}]{CIRetal2007}
{Cirtain}, J.~W., {Golub}, L., {Lundquist}, L., {et~al.} 2007,
Science, 318,
  1580

\bibitem[{{Curdt} \& {Tian}(2011)}]{CURTIA2011}
{Curdt}, W. \& {Tian}, H. 2011, \aap, 532, L9+

\bibitem[{{De Moortel}(2009)}]{DEM2009}
{De Moortel}, I. 2009, Space Science Reviews, 149, 65

\bibitem[{{De Pontieu} {et~al.}(2004){De Pontieu}, {Erd{\'e}lyi}, \&
  {James}}]{DEPetal2004}
{De Pontieu}, B., {Erd{\'e}lyi}, R., \& {James}, S.~P. 2004, \nat,
430, 536

\bibitem[{{De Pontieu} {et~al.}(2007{\natexlab{a}}){De Pontieu}, {McIntosh},
  {Hansteen}, {Carlsson}, {Schrijver}, {Tarbell}, {Title}, {Shine}, {Suematsu},
  {Tsuneta}, {Katsukawa}, {Ichimoto}, {Shimizu}, \& {Nagata}}]{DEPetal2007a}
{De Pontieu}, B., {McIntosh}, S., {Hansteen}, V.~H., {et~al.}
  2007{\natexlab{a}}, \pasj, 59, 655

\bibitem[{{De Pontieu} \& {McIntosh}(2010)}]{DEPMCI2010}
{De Pontieu}, B. \& {McIntosh}, S.~W. 2010, \apj, 722, 1013

\bibitem[{{De Pontieu} {et~al.}(2011){De Pontieu}, {McIntosh}, {Carlsson},
  {Hansteen}, {Tarbell}, {Boerner}, {Martinez-Sykora}, {Schrijver}, \&
  {Title}}]{DEPetal2011}
{De Pontieu}, B., {McIntosh}, S.~W., {Carlsson}, M., {et~al.} 2011,
Science,
  331, 55

\bibitem[{{De Pontieu} {et~al.}(2007{\natexlab{b}}){De Pontieu}, {McIntosh},
  {Carlsson}, {Hansteen}, {Tarbell}, {Schrijver}, {Title}, {Shine}, {Tsuneta},
  {Katsukawa}, {Ichimoto}, {Suematsu}, {Shimizu}, \& {Nagata}}]{DEPetal2007}
{De Pontieu}, B., {McIntosh}, S.~W., {Carlsson}, M., {et~al.}
  2007{\natexlab{b}}, Science, 318, 1574

\bibitem[{{De Pontieu} {et~al.}(2009){De Pontieu}, {McIntosh}, {Hansteen}, \&
  {Schrijver}}]{DEPetal2009}
{De Pontieu}, B., {McIntosh}, S.~W., {Hansteen}, V.~H., \&
{Schrijver}, C.~J.
  2009, \apjl, 701, L1

\bibitem[{{Erd{\'e}lyi}(2006{\natexlab{a}})}]{ERD2006b}
{Erd{\'e}lyi}, R. 2006{\natexlab{a}}, Royal Society of London
Philosophical
  Transactions Series A, 364, 351

\bibitem[{{Erd{\'e}lyi}(2006{\natexlab{b}})}]{ERD2006}
{Erd{\'e}lyi}, R. 2006{\natexlab{b}}, in Proceedings of SOHO 18/GONG
2006/HELAS
  I, Beyond the spherical Sun, ed. K.~{Fletcher} \& M.~{Thompson}, ESA SP-624

\bibitem[{{Hansteen} {et~al.}(2006){Hansteen}, {De Pontieu}, {Rouppe van der
  Voort}, {van Noort}, \& {Carlsson}}]{HANetal2006}
{Hansteen}, V.~H., {De Pontieu}, B., {Rouppe van der Voort}, L.,
{van Noort},
  M., \& {Carlsson}, M. 2006, \apjl, 647, L73

\bibitem[{{Heggland} {et~al.}(2007){Heggland}, {De Pontieu}, \&
  {Hansteen}}]{HEGetal2007}
{Heggland}, L., {De Pontieu}, B., \& {Hansteen}, V.~H. 2007, \apj,
666, 1277

\bibitem[{{Heggland} {et~al.}(2009){Heggland}, {De Pontieu}, \&
  {Hansteen}}]{HEGetal2009}
{Heggland}, L., {De Pontieu}, B., \& {Hansteen}, V.~H. 2009, \apj,
702, 1

\bibitem[{{Hollweg} {et~al.}(1982){Hollweg}, {Jackson}, \&
  {Galloway}}]{HOLetal1982}
{Hollweg}, J.~V., {Jackson}, S., \& {Galloway}, D. 1982, \solphys,
75, 35

\bibitem[{{Huang} {et~al.}(1998){Huang}, {Shen}, {Long}, {Wu}, {Shih}, {Zheng},
  {Yen}, {Tung}, \& {Liu}}]{HUAetal1998}
{Huang}, N.~E., {Shen}, Z., {Long}, S.~R., {et~al.} 1998, Royal
Society of
  London Proceedings Series A, 454, 903

\bibitem[{{Jess} {et~al.}(2009){Jess}, {Mathioudakis}, {Erd{\'e}lyi},
  {Crockett}, {Keenan}, \& {Christian}}]{JESetal2009}
{Jess}, D.~B., {Mathioudakis}, M., {Erd{\'e}lyi}, R., {et~al.} 2009,
Science,
  323, 1582

\bibitem[{{Kamio} {et~al.}(2010){Kamio}, {Curdt}, {Teriaca}, {Inhester}, \&
  {Solanki}}]{KAMetal2010}
{Kamio}, S., {Curdt}, W., {Teriaca}, L., {Inhester}, B., \&
{Solanki}, S.~K.
  2010, \aap, 510, L1+

\bibitem[{{Kuridze} {et~al.}(2012){Kuridze}, {Morton}, {Erd\'elyi}, {Dorrian},
  {Mathioudakis}, {Jess}, \& {Keenan}}]{KURetal2012}
{Kuridze}, D., {Morton}, R.~J., {Erd\'elyi}, R., {et~al.}
2012-Accepted, \apjl

\bibitem[{{Lemen} {et~al.}(2011){Lemen}, {Title}, {Akin}, {Boerner}, {Chou},
  {Drake}, {Duncan}, {Edwards}, {Friedlaender}, {Heyman}, {Hurlburt}, {Katz},
  {Kushner}, {Levay}, {Lindgren}, {Mathur}, {McFeaters}, {Mitchell}, {Rehse},
  {Schrijver}, {Springer}, {Stern}, {Tarbell}, {Wuelser}, {Wolfson}, {Yanari},
  {Bookbinder}, {Cheimets}, {Caldwell}, {Deluca}, {Gates}, {Golub}, {Park},
  {Podgorski}, {Bush}, {Scherrer}, {Gummin}, {Smith}, {Auker}, {Jerram},
  {Pool}, {Soufli}, {Windt}, {Beardsley}, {Clapp}, {Lang}, \&
  {Waltham}}]{LEMetal2011}
{Lemen}, J.~R., {Title}, A.~M., {Akin}, D.~J., {et~al.} 2011,
\solphys, 115

\bibitem[{{Liu} {et~al.}(2009){Liu}, {Berger}, {Title}, \&
  {Tarbell}}]{LIUetal2009}
{Liu}, W., {Berger}, T.~E., {Title}, A.~M., \& {Tarbell}, T.~D.
2009, \apjl,
  707, L37

\bibitem[{{Liu} {et~al.}(2011){Liu}, {Berger}, {Title}, {Tarbell}, \&
  {Low}}]{LIUetal2011}
{Liu}, W., {Berger}, T.~E., {Title}, A.~M., {Tarbell}, T.~D., \&
{Low}, B.~C.
  2011, \apj, 728, 103

\bibitem[{{Mart{\'{\i}}nez-Sykora} {et~al.}(2011){Mart{\'{\i}}nez-Sykora},
  {Hansteen}, \& {Moreno-Insertis}}]{MARetal2011}
{Mart{\'{\i}}nez-Sykora}, J., {Hansteen}, V., \& {Moreno-Insertis},
F. 2011,
  \apj, 736, 9

\bibitem[{{McLaughlin} {et~al.}(2010){McLaughlin}, {Hood}, \& {de
  Moortel}}]{MCLetal2010}
{McLaughlin}, J.~A., {Hood}, A.~W., \& {de Moortel}, I. 2010, \ssr,
62

\bibitem[{{Moore} {et~al.}(2010){Moore}, {Cirtain}, {Sterling}, \&
  {Falconer}}]{MOOetal2010}
{Moore}, R.~L., {Cirtain}, J.~W., {Sterling}, A.~C., \& {Falconer},
D.~A. 2010,
  \apj, 720, 757

\bibitem[{{Moore} {et~al.}(2011){Moore}, {Sterling}, {Cirtain}, \&
  {Falconer}}]{MOOetal2011}
{Moore}, R.~L., {Sterling}, A.~C., {Cirtain}, J.~W., \& {Falconer},
D.~A. 2011,
  \apjl, 731, L18+

\bibitem[{{Morton} {et~al.}(2011{\natexlab{a}}){Morton}, {Erd{\'e}lyi}, {Jess},
  \& {Mathioudakis}}]{MORetal2011}
{Morton}, R.~J., {Erd{\'e}lyi}, R., {Jess}, D.~B., \&
{Mathioudakis}, M.
  2011{\natexlab{a}}, \apjl, 729, L18

\bibitem[{{Morton} {et~al.}(2011{\natexlab{b}}){Morton}, {Verth}, {McLaughlin},
  \& {Erd{\'e}lyi}}]{MORetal2011b}
{Morton}, R.~J., {Verth}, G., {McLaughlin}, J.~A., \& {Erd{\'e}lyi},
R.
  2011{\natexlab{b}}, \apj

\bibitem[{{Murawski} \& {Zaqarashvili}(2010)}]{MURZAQ2010}
{Murawski}, K. \& {Zaqarashvili}, T.~V. 2010, \aap, 519, A8

\bibitem[{{Newton}(1934)}]{NEW1934}
{Newton}, H.~W. 1934, \mnras, 94, 472

\bibitem[{{Pariat} {et~al.}(2009){Pariat}, {Antiochos}, \&
  {DeVore}}]{PARetal2009}
{Pariat}, E., {Antiochos}, S.~K., \& {DeVore}, C.~R. 2009, \apj,
691, 61

\bibitem[{{Patsourakos} {et~al.}(2008){Patsourakos}, {Pariat}, {Vourlidas},
  {Antiochos}, \& {Wuelser}}]{PATetal2008}
{Patsourakos}, S., {Pariat}, E., {Vourlidas}, A., {Antiochos},
S.~K., \&
  {Wuelser}, J.~P. 2008, \apjl, 680, L73

\bibitem[{{Pike} \& {Mason}(1998)}]{PIKMAS1998}
{Pike}, C.~D. \& {Mason}, H.~E. 1998, \solphys, 182, 333

\bibitem[{{Roberts} {et~al.}(1984){Roberts}, {Edwin}, \& {Benz}}]{ROBetal1984}
{Roberts}, B., {Edwin}, P.~M., \& {Benz}, A.~O. 1984, \apj, 279, 857

\bibitem[{{Ruderman} \& {Erd{\'e}lyi}(2009)}]{RUDERD2009}
{Ruderman}, M.~S. \& {Erd{\'e}lyi}, R. 2009, Space Science Reviews,
149, 199

\bibitem[{{Schrijver} {et~al.}(1998){Schrijver}, {Title}, {Harvey}, {Sheeley},
  {Wang}, {van den Oord}, {Shine}, {Tarbell}, \& {Hurlburt}}]{SCHetal1998}
{Schrijver}, C.~J., {Title}, A.~M., {Harvey}, K.~L., {et~al.} 1998,
\nat, 394,
  152

\bibitem[{{Scullion} {et~al.}(2011){Scullion}, {Erd\'elyi}, {Fedun}, \&
  {Doyle}}]{SCUetal2011}
{Scullion}, E., {Erd\'elyi}, R., {Fedun}, V., \& {Doyle}, J. 2011,
Submitted

\bibitem[{{Secchi}(1877)}]{SEC1877}
{Secchi}, P.~A. 1877, La Soleil, Vol. 2, Chapter II (Paris:
Gauthier-Villars)

\bibitem[{{Shibata} {et~al.}(1992){Shibata}, {Ishido}, {Acton}, {Strong},
  {Hirayama}, {Uchida}, {McAllister}, {Matsumoto}, {Tsuneta}, {Shimizu},
  {Hara}, {Sakurai}, {Ichimoto}, {Nishino}, \& {Ogawara}}]{SHIetal1992}
{Shibata}, K., {Ishido}, Y., {Acton}, L.~W., {et~al.} 1992, \pasj,
44, L173

\bibitem[{{Shibata} {et~al.}(2007){Shibata}, {Nakamura}, {Matsumoto}, {Otsuji},
  {Okamoto}, {Nishizuka}, {Kawate}, {Watanabe}, {Nagata}, {UeNo}, {Kitai},
  {Nozawa}, {Tsuneta}, {Suematsu}, {Ichimoto}, {Shimizu}, {Katsukawa},
  {Tarbell}, {Berger}, {Lites}, {Shine}, \& {Title}}]{SHIetal2007}
{Shibata}, K., {Nakamura}, T., {Matsumoto}, T., {et~al.} 2007,
Science, 318,
  1591

\bibitem[{{Shibata} \& {Uchida}(1985)}]{SHIUCH1985}
{Shibata}, K. \& {Uchida}, Y. 1985, \pasj, 37, 31

\bibitem[{{Shimojo} {et~al.}(1996){Shimojo}, {Hashimoto}, {Shibata},
  {Hirayama}, {Hudson}, \& {Acton}}]{SHIMetal1996}
{Shimojo}, M., {Hashimoto}, S., {Shibata}, K., {et~al.} 1996, \pasj,
48, 123

\bibitem[{{Srivastava} \& {Murawski}(2011)}]{SRIMUR2011}
{Srivastava}, A.~K. \& {Murawski}, K. 2011, \aap, 534, A62

\bibitem[{{Sterling} {et~al.}(2010){Sterling}, {Harra}, \&
  {Moore}}]{STEetal2010}
{Sterling}, A.~C., {Harra}, L.~K., \& {Moore}, R.~L. 2010, \apj,
722, 1644

\bibitem[{{Taroyan} \& {Erd{\'e}lyi}(2009)}]{TARERD2009}
{Taroyan}, Y. \& {Erd{\'e}lyi}, R. 2009, Space Science Reviews, 24

\bibitem[{{Terradas} {et~al.}(2004){Terradas}, {Oliver}, \&
  {Ballester}}]{TERetal2004}
{Terradas}, J., {Oliver}, R., \& {Ballester}, J.~L. 2004, \apj, 614,
435

\bibitem[{{Tian} {et~al.}(2011){Tian}, {McIntosh}, \& {De
  Pontieu}}]{TIAetal2011}
{Tian}, H., {McIntosh}, S.~W., \& {De Pontieu}, B. 2011, \apjl, 727,
L37+

\bibitem[{{Uchida}(1970)}]{UCHIDA1970}
{Uchida}, Y. 1970, PASJ, 22, 341

\bibitem[{{Vasheghani Farahani} {et~al.}(2009){Vasheghani Farahani}, {Van
  Doorsselaere}, {Verwichte}, \& {Nakariakov}}]{VAGetal2009}
{Vasheghani Farahani}, S., {Van Doorsselaere}, T., {Verwichte}, E.,
\&
  {Nakariakov}, V.~M. 2009, \aap, 498, L29

\bibitem[{{Verwichte} {et~al.}(2010){Verwichte}, {Marsh}, {Foullon}, {Van
  Doorsselaere}, {De Moortel}, {Hood}, \& {Nakariakov}}]{VERetal2010}
{Verwichte}, E., {Marsh}, M., {Foullon}, C., {et~al.} 2010, \apjl,
724, L194

\bibitem[{{Wang}(2011)}]{WAN2011}
{Wang}, T. 2011, \ssr, 16

\bibitem[{{Wang} {et~al.}(2011){Wang}, {Ofman}, \& {Davila}}]{WANetal2011}
{Wang}, T., {Ofman}, L., \& {Davila}, J.~M. 2011, ArXiv e-prints

\bibitem[{{Yokoyama} \& {Shibata}(1995)}]{YOKSHI1995}
{Yokoyama}, T. \& {Shibata}, K. 1995, \nat, 375, 42

\bibitem[{{Yokoyama} \& {Shibata}(1996)}]{YOKSHI1996}
{Yokoyama}, T. \& {Shibata}, K. 1996, \pasj, 48, 353

\bibitem[{{Zaqarashvili} \& {Erd{\'e}lyi}(2009)}]{ZAQERD2009}
{Zaqarashvili}, T.~V. \& {Erd{\'e}lyi}, R. 2009, \ssr, 149, 355

\end{thebibliography}

\end{document}